\documentclass[10pt,superscriptaddress,twocolumn,amsmath,amssymb,aps,prl,showpacs,groupedaddress]{revtex4-1}
\usepackage{mathrsfs}
\usepackage{graphicx}
\usepackage{dcolumn}
\usepackage{bm}

\usepackage{graphicx}
\usepackage{epstopdf}

\usepackage{amsmath,amssymb,mathrsfs}
\usepackage{braket}
\usepackage{hyperref}
\hypersetup{
    colorlinks=true,
    linkcolor=blue,
    filecolor=blue,
    urlcolor=blue,
    citecolor=magenta
    }

\begin{document}


\title{Fragmented Condensate in a Two-Component Bose Gas with $p$-wave Interactions}


\author{Mingyang Liu}
\affiliation{Department of Physics and Hong Kong Institute of Quantum Science and Technology, The University of Hong Kong, Hong Kong, China}
\author{Shizhong Zhang}
\affiliation{Department of Physics and Hong Kong Institute of Quantum Science and Technology, The University of Hong Kong, Hong Kong, China}


\date{\today}

\begin{abstract} 
In this Letter, we discuss the effects of $p$-wave attractive interaction in a spin-$1/2$ Bose gas. With a repulsive $s$-wave background interaction, we show that for weak $p$-wave attraction, one obtains a standard Bose-Einstein condensate at zero momentum with spins fully polarized. Upon increasing the $p$-wave attraction, a fragmented condensate state with singlet pair formation and $p$-wave correlation emerges. We determine the transition point and investigate the properties of the fragmented condensate using an ansatz wave function. We construct the relevant Gross-Pitaevskii equations for the fragmented condensate and show that the sound velocities are anisotropic and may vanish in specific directions. Based on the many-body wave function, we also discuss the low-energy spin excitations of the system.
\end{abstract}
\maketitle

{\it Introduction}.--The inclusion of spin degrees of freedom into Bose condensate has led to a variety of new quantum states and excitations that have no analog in the traditional condensate such as liquid Helium four~\cite{SKMU}. They offer an ideal platform to investigate the interplay between superfluidity and magnetism. In particular, many exotic topological excitations, such as skymions~\cite{Choi2012} and Mermin-Ho textures~\cite{Leslie2009, Jae-yoon2012}, are observed experimentally. Coherent spin mixing dynamics, unique to spinor quantum gases, have also been observed in both the ferromagnetic and antiferromagnetic spinor condensate, enabling the determination of spin-dependent interactions and the observation of spin-mixing resonances~\cite{Chang2005,Kronjager2006,Black2007,Jung2009}.

The situation for spin-$1/2$ boson is quite different. This is understandable since without spin-orbit coupling, the standard condensation would occur in a unique single particle state that is a direct product of spatial and spin wave functions; being spin-$1/2$, all bosons are then polarized along a certain direction $\hat{n}$ in the spin space, thus leading to a ferromagnetic condensate with maximal total spin projection given by $N\hbar/2$, where $N$ is the total number of bosons under consideration. Condensate with more exotic structure can be realized if one imposes the constraint that the total spin of the system must be close to zero which is appropriate in the high temperature regime, and this has led to the discussion of a metastable fragmented condensate, the so-called KSA state~\cite{KS,AL,Leggettbook}, but it has so far not been realized experimentally.

In this Letter, we show that by including a sizeable $p$-wave interaction between spin-$1/2$ bosons, it is possible to realize the analog of the KSA state as the ground state. In addition, it is possible to tune the $p$-wave interaction so that one can effect a transition from the standard condensate at zero momentum to the fragmented state that involves two opposite momentum. The fragmented condensate with its broken symmetry in the orbital space leads to novel density and spin excitations. In addition, our investigation also highlights the importance of the $p$-wave effective range in determining the structure of the condensate. 

{\it The Model.}-- Let us now consider a spin-$1/2$ boson system, such as $^{87}$Rb in their lowest two hyperfine-Zeeman states and assume that the background $s$-wave interaction is isotropic and is characterized by a single $s$-wave scattering length $a_0$. In addition, we also include a $p$-wave interaction between the two spin species. Due to Bose symmetry, it is necessary that the two bosons should form a spin singlet in the $p$-wave channel. The effective $p$-wave interaction $U^{\ell =1}_{{\bf k},{\bf k}'}$ is characterised by the low-energy scattering $T$-matrix, which, when expanded to include the effective range, takes the form ($\hbar=1$)
\begin{equation}
U^{\ell =1}_{{\bf k},{\bf k}'}=T^{\ell =1}_{{\bf k},{\bf k}'}=\frac{12\pi}{m}\left(a_1+\frac{1}{2}a_1^2r_ek^2+\cdots\right){\bf k}\cdot{\bf k}',
\end{equation}
where $a_1$ is the $p$-wave scattering volume and $r_e$ is the $p$-wave effective range. We note that $p$-wave interaction vanishes when either $|{\bf k}|$ or $|{\bf k}'|$ approaches zero and is in general anisotropic in the momentum space. Together with the $s$-wave scattering $U^{\ell =0}_{{\bf k},{\bf k}'}\equiv U_0=4\pi a_0/m$ where $a_0$ is the $s$-wave scattering length, the effective interaction in our model can be written as $U_{{\bf k},{\bf k}'}=U^{\ell =0}_{{\bf k},{\bf k}'}+U^{\ell =1}_{{\bf k},{\bf k}'}$. We will neglect higher partial wave contribution and furthermore consider attractive $p$-wave interaction ($a_1<0$) in the following discussion. The Hamiltonian of the system can then be written as
\begin{equation}
\begin{split}
H & =\sum_{\mathbf{k},\sigma}\frac{k^{2}}{2m}a_{\mathbf{k}\sigma}^{\dagger}a_{\mathbf{k}\sigma}\\
&+\frac{1}{V}\sum_{\mathbf{P},\mathbf{k}_{1},\mathbf{k}_{2}}(U_{\mathbf{k}_{1},\mathbf{k}_{2}}^{\ell=1}+U_0) a_{\mathbf{\frac{P}{2}+k_{1}}\uparrow}^{\dagger}a_{\mathbf{\frac{P}{2}-k_{1}}\downarrow}^{\dagger}a_{\mathbf{\frac{P}{2}-k}_{2}\downarrow}a_{\mathbf{\frac{P}{2}+k}_{2}\uparrow}\\
&+\frac{U_0}{2V}\sum_{\mathbf{P},\mathbf{k}_{1},\mathbf{k}_{2},\sigma}a_{\mathbf{\frac{P}{2}+k_{1}}\sigma}^{\dagger}a_{\mathbf{\frac{P}{2}-k_{1}}\sigma}^{\dagger} a_{\mathbf{\frac{P}{2}-k_{2}}\sigma}a_{\mathbf{\frac{P}{2}+k_{2}}\sigma}\label{Hamiltonian}
\end{split}
\end{equation}
Here $a_{\mathbf{k}\sigma}^{\dagger}$ is the creation operator for bosons with momentum $\mathbf{k}$ and spin $\sigma$. The second term contains interaction in both the spin singlet channel for which $U_{\mathbf{k}_{1},\mathbf{k}_{2}}^{\ell=1}$ is operative and the triplet channel for which $U_0$ makes nonzero contributions. The last term describes the $s$-wave interaction between identical bosons in the triplet channel.  

To determine the structure of the ground state, we first observe that the conventional condensate with all bosons occupying the zero-momentum state would experience only $s$-wave interaction with energy density given by $n^2U_0/2$, where $n$ is the average density. To exploit the attractive $p$-wave interaction, it is not only necessary that the wave function develop non-zero momentum component, but that it must be correlated in such a way to give rise to a non-zero singlet component in the two-body correlation. Furthermore, due to Galilean invariance, the ground state must carry zero momentum, this suggests that the singlet component of the wave function must be of the form $(a^\dagger_{{\bf p},\uparrow}a^\dagger_{-{\bf p},\downarrow}-a^\dagger_{-{\bf p},\uparrow}a^\dagger_{{\bf p},\downarrow})\ket{0}$ where $\ket{0}$ is the vacuum state. In fact, were we to consider a two-body state of the above form within a finite box of volume $V$, the kinetic energy would be simply $p^2/m$, and after adding the interaction energy, the total energy is given by $(p^2/m)(1+48\pi a_1/V)+24\pi a_1^2r_e p^4/(mV)$. It is then clear that when $1+48\pi a_1/V<0$, the optimal value of $p$ is finite for $r_e>0$ and the ground state energy is negative. For $r_e<0$, the higher order terms in the effective range expansion must be taken into account to obtain a finite value of $p$. In free space, one generally has $r_e<0$ due to causality bound~\cite{HammerLee}, but this can be tuned by imposing a weak optical lattice, as discussed in the supplementary material~\cite{supp}. In the following, we shall assume $r_e>0$ for our purpose. We note that for the two-body state described above, there is no $s$-wave interaction. To incorporate this singlet correlation for a system of $N$-bosons, we adopt the following ansatz, assuming that $N$ is even,
\begin{equation}
\ket{\mbox{G}}=\mathcal{Z}\left(a^\dagger_{{\bf p},\uparrow}a^\dagger_{-{\bf p},\downarrow}-a^\dagger_{-{\bf p},\uparrow}a^\dagger_{{\bf p},\downarrow}\right)^{N/2}\ket{0}\label{ansatz1}
\end{equation}
with ${\bf p}$ to be chosen by minimizing the energy expectation value $E({\bf p})\equiv \bra{G} H\ket{G}$. $\mathcal{Z}=[(N/2)!(N/2+1)!]^{-1/2}$ is the normalization constant. We note that for the many-body wave function eq. (\ref{ansatz1}), the $s$-wave interaction is non-zero, unlike in the two-body case. It can also be written as
\begin{equation}
\ket{\mbox{G}}=\frac{\sqrt{N/2}}{(N/2)!}\int\frac{d\Omega}{4\pi}(a_{\bf p}^\dagger(\Omega)a_{-\bf p}^\dagger(-\Omega))^{N/2}\ket{0}\label{ansatz2}
\end{equation}
where $a_{\bf p}^\dagger(\Omega)$ creates bosons with momentum ${\bf p}$ and spin ${\bf s}$ along the direction $\hat{d}$ specified by the solid angle $\Omega$, that satisfies ${\bf s}\cdot\hat{d}=1/2$.

Before evaluating $E({\bf p})$ explicitly, several remarks are in order. First, if we denote the total spin of the system as ${\bf S}=\sum_i{\bf s}_i$ with ${\bf s}_i$ the spin of each boson, then eq.(\ref{ansatz1}) is an eigenstate of ${\bf S}$ with ${\bf S}^2=0$. Namely, it is a many-body spin singlet and the $SU(2)$ symmetry in spin space is  unbroken. This is in stark contrast with the conventional condensate for which ${\bf S}^2=N/2(N/2+1)$, being a maximally polarized state. In particular, we note that eq.(\ref{ansatz1}) is an eigenstates of $N_{\uparrow}=a^\dagger_{{\bf p},\uparrow}a_{{\bf p},\uparrow}+a^\dagger_{-{\bf p},\uparrow}a_{-{\bf p},\uparrow}$ with eigenvalue $N/2$. The same for $N_{\downarrow}$. Thus there is no fluctuation of atom numbers. This is different from the so-called LPB state~\cite{LPB} which exhibits anomalous number fluctuations~\cite{HoYip}. Secondly, the state $\ket{\mbox{G}}$ carries zero charge or spin current, different from the mean field state proposed for the mixture of two bosonic species in the pioneering work before~\cite{RCPRL, CRPRB, Li2019}.  Thirdly, by choosing a particular value of ${\bf p}$, the $SO(3)$ symmetry in orbital space is broken to $O(2)\times Z_2$ (due to the symmetry between ${-{\bf p}}$ and ${\bf p}$). As we shall show later, this has important consequences for the elementary excitations of the system. Finally, as discussed previously, eq.(\ref{ansatz1}) describes a fragmented condensate where four single particle states, $\ket{{\bf p}_0\uparrow}$, $\ket{{\bf p}_0\downarrow}$,$\ket{-{\bf p}_0\uparrow}$,$\ket{-{\bf p}_0\downarrow}$, are macroscopically occupied, each with eigenvalue $N/4$. For a more general discussion of fragmented condensate, see Refs.~\cite{Castin2001, Mueller2006}.

{\it Ground State Properties.}--Using eqs.(\ref{Hamiltonian}) and (\ref{ansatz1}), we find that $E({\bf p})$ is given by
\begin{equation}
E({\bf p})=\frac{{\bf p}^2}{2m}N+\frac{N}{2V}\left(\frac{N}{2}+1\right)U_{\mathbf{k}_{1},\mathbf{k}_{2}}^{\ell=1}+\frac{N}{V}\left(\frac{N}{2}-1\right)U_0.\label{energy_finite_N}
\end{equation}
In the thermodynamic limit, using the explicit form for $U_{\mathbf{k}_{1},\mathbf{k}_{2}}^{\ell=1}$, we can write the energy density as
\begin{equation}
\frac{E({\bf p})}{V}=\frac{1}{2}n^2 U_0+\frac{n}{2m}(1+6\pi na_1){\bf p}^2+\frac{3\pi}{2m}n^2a_1^2r_e {\bf p}^4,\label{energy_single_p}
\end{equation}
where $n=N/V$ is the average density. The first term is simply the contribution from $s$-wave scattering. The second and third terms have the same structure as the two-body case but now with the effective volume replaced by $\sim 1/n$. In this case, when
\begin{equation}
1+6\pi na_1<0,\label{cond1}
\end{equation}
a finite value of ${\bf p}={\bf p}_0$ is obtained and is given by 
\begin{equation}
|{\bf p}_0|=\sqrt{-\frac{1+6\pi n a_1}{6\pi a_1^2 r_e n}}.\label{valuep_0}
\end{equation}
For $na_1\sim 1$, $|{\bf p}_0|\sim (na_1^2r_e)^{-1/2}$ that depends on interaction and density, different from the minimum value of $|{\bf p}|$ determined by the single particle physics~\cite{AL}, and is similar to the exactly solvable case considered in Ref.~\cite{SLH}. On the other hand, when $1+6\pi na_1>0$, the minimal energy occurs at ${\bf p}_0=0$. This corresponds to a standard condensate when all particle condenses in the zero momentum state and $p$-wave interaction vanishes. Thus, by tuning the $p$-wave attraction, we can go from a standard condensate to a fragmented one that is described by eq.(\ref{ansatz1}). In addition, it is necessary to have a positive compressibility $\kappa$ to avoid mechanical collapse of the condensate. Using $\kappa^{-1}=n^2\partial^2 (E({\bf p}_0)/V)/\partial n^2>0$, we obtain the constraint on $a_0$:
\begin{equation}
a_0>\frac{3}{4 r_e}.\label{cond2}
\end{equation}
Eqs.(\ref{cond1}) and (\ref{cond2}) together give the conditions under which the ansatz state (\ref{ansatz1}) is stable. 

The wave function eq.(\ref{ansatz1}) is analogous to the fragmented state proposed in Refs.~\cite{KS,AL} for spin-$1/2$ bosons for which the two orbital states correspond to the lowest and first excited states of the single particle Hamiltonian, and is stablized because of spin conservation; In our case, it is induced by a non-zero attractive $p$-wave interaction which selects two degenerate momentum states $\ket{{\bf p}}$ and $\ket{-{\bf p}}$ and stablized by a repulsive $s$-wave interaction. In the presence of residue magnetic field that breaks the spin $SU(2)$ symmetry, the actual ground state might correspond a particular choice of $\Omega$ in Eq.(\ref{ansatz2}), thus generating the standard symmetry breaking state with antiferromagnetic corrections.

{\it Density Fluctuations.}-- Below, we discuss the possible sound mode in the condensate with the structure given by eq.(\ref{ansatz1}). In order to generate sound excitations, it is necessary that the corresponding phonon wave function features momentum component other than $\pm {\bf p}_0$. To do that, let us define two general orthogonal single particle states labeled by $0$ and $1$,
\begin{equation}
a_{0\sigma}=\sum_{\bf k}c_{\bf k}a_{{\bf k}\sigma},~~~~a_{1\sigma}=\sum_{\bf k}d_{\bf k}a_{{\bf k}\sigma}.
\end{equation}
Note that the coefficient $c_{\bf k}$ and $d_{\bf k}$ do not depend on the spin index since we are considering density fluctuations. We also need to impose the normalization conditions $\sum_{\bf k}|c_{\bf k}|^2=\sum_{\bf k}|d_{\bf k}|^2=1$ and orthogonal condition $\sum_{\bf k}c^*_{\bf k}d_{\bf k}=\sum_{\bf k}c_{\bf k}d^*_{\bf k}=0$. The general phonon wave function is still of the form given in eq.(\ref{ansatz1}), but now with the single particle states $\pm {\bf p}_0$ replaced by $0,1$. Thus
\begin{equation}
\ket{\mbox{phonon}}=\mathcal{Z}\left(a^\dagger_{0,\uparrow}a^\dagger_{1,\downarrow}-a^\dagger_{1,\uparrow}a^\dagger_{0,\downarrow}\right)^{N/2}\ket{0}\label{phonon}.
\end{equation}
We note that the spin structure of eq.(\ref{phonon}) is exactly the same as that of eq.(\ref{ansatz1}). It is straightforward to evaluate the average energy of the state (\ref{phonon}) and the final result is
\begin{equation}
\begin{split}
E &=\frac{N}{2}\sum_{\bf k}\frac{{\bf k}^2}{2m}(|c_{\bf k}|^2+|d_{\bf k}|^2)\\
   &+\frac{N}{2V}\left(\frac{N}{2}+1\right)\sum_{{\bf P},{\bf k}_1, {\bf k}_2}U_{\mathbf{k}_{1},\mathbf{k}_{2}}^{\ell=1}c^*_{\frac{\bf P}{2}+{\bf k}_2}d^*_{\frac{\bf P}{2}-{\bf k}_2}d_{\frac{\bf P}{2}-{\bf k}_1}c_{\frac{\bf P}{2}+{\bf k}_1}\\
   &+\frac{N}{2V}\left(\frac{N}{2}-1\right)\sum_{{\bf P},{\bf k}_1, {\bf k}_2}U_0c^*_{\frac{\bf P}{2}+{\bf k}_2}d^*_{\frac{\bf P}{2}-{\bf k}_2}d_{\frac{\bf P}{2}-{\bf k}_1}c_{\frac{\bf P}{2}+{\bf k}_1}\\
   &+\frac{N}{4V}\left(\frac{N}{2}-1\right)\sum_{{\bf P},{\bf k}_1, {\bf k}_2}U_0(c^*_{\frac{\bf P}{2}+{\bf k}_2}c^*_{\frac{\bf P}{2}-{\bf k}_2}c_{\frac{\bf P}{2}-{\bf k}_1}c_{\frac{\bf P}{2}+{\bf k}_1}\\
   &+d^*_{\frac{\bf P}{2}+{\bf k}_2}d^*_{\frac{\bf P}{2}-{\bf k}_2}d_{\frac{\bf P}{2}-{\bf k}_1}d_{\frac{\bf P}{2}+{\bf k}_1}).\label{energy}
\end{split}
\end{equation}
We can then use the method of Lagrangian multipliers to minimize the energy (\ref{energy}) subjected to the normalization and orthogonalisation constraints. Introducing the multipliers $\mu_0,\mu_1$ and $\lambda,\lambda^*$, we then need to find the minimum of $F\equiv E-\frac{N\mu_0}{2}\sum_{\bf k}|c_{\bf k}|^2-\frac{N\mu_1}{2}\sum_{\bf k}|d_{\bf k}|^2-\frac{N\lambda^*}{2}\sum_{\bf k}d^*_{\bf k}c_{\bf k}-\frac{N\lambda}{2}\sum_{\bf k}d_{\bf k}c^*_{\bf k}$ with respect to the variation of the complex amplitudes $c_{\bf k}$ and $d_{\bf k}$. This leaves us with the following two coupled Gross-Pitaevskii equations
\begin{align}\nonumber
&\frac{{\bf k}^2}{2m}c_{\bf k}+\frac{n}{2}\sum_{{\bf k}_1,{\bf k}_2}(U_{\mathbf{k}_{1},\mathbf{k}_{2}}^{\ell=1}+U_0)d^*_{{\bf k}-2{\bf k}_2}d_{{\bf k}-{\bf k}_1-{\bf k}_2}c_{{\bf k}+{\bf k}_1-{\bf k}_2}\\
&+U_0c^*_{{\bf k}-2{\bf k}_1}c_{{\bf k}-{\bf k}_1-{\bf k}_2}c_{{\bf k}+{\bf k}_1-{\bf k}_2}=\mu_0 c_{\bf k}+\lambda d_{\bf k}\\\nonumber
&\frac{{\bf k}^2}{2m}d_{\bf k}+\frac{n}{2}\sum_{{\bf k}_1,{\bf k}_2}(U_{\mathbf{k}_{1},\mathbf{k}_{2}}^{\ell=1}+U_0)c^*_{{\bf k}-2{\bf k}_2}c_{{\bf k}-{\bf k}_1-{\bf k}_2}d_{{\bf k}+{\bf k}_1-{\bf k}_2}\\
&+U_0d^*_{{\bf k}-2{\bf k}_1}d_{{\bf k}-{\bf k}_1-{\bf k}_2}d_{{\bf k}+{\bf k}_1-{\bf k}_2}=\mu_0 c_{\bf k}+\lambda d_{\bf k}
\end{align}
These two equations describes the density fluctuations of the fragmented condensate, analogous to the standard GP equation. One can check that 
\begin{equation}
c_{\bf k}=\delta_{{\bf k},{\bf p}}~\mbox{and}~d_{\bf k}=\delta_{{\bf k},-{\bf p}}\label{cdsolution}
\end{equation}
is a solution of the coupled equations with
\begin{align}
\mu_0 &=\mu_1=\frac{{\bf p}^2}{2m}+\frac{n}{2}U_{\mathbf{p},\mathbf{p}}^{\ell=1}+nU_0\\
\lambda &=\lambda^*=0
\end{align}
and substitution of eq.(\ref{cdsolution}) to eq.(\ref{energy}) reproduces eq.(\ref{energy_single_p}), as it should. 

\begin{figure}[h]
\includegraphics[scale=0.36]{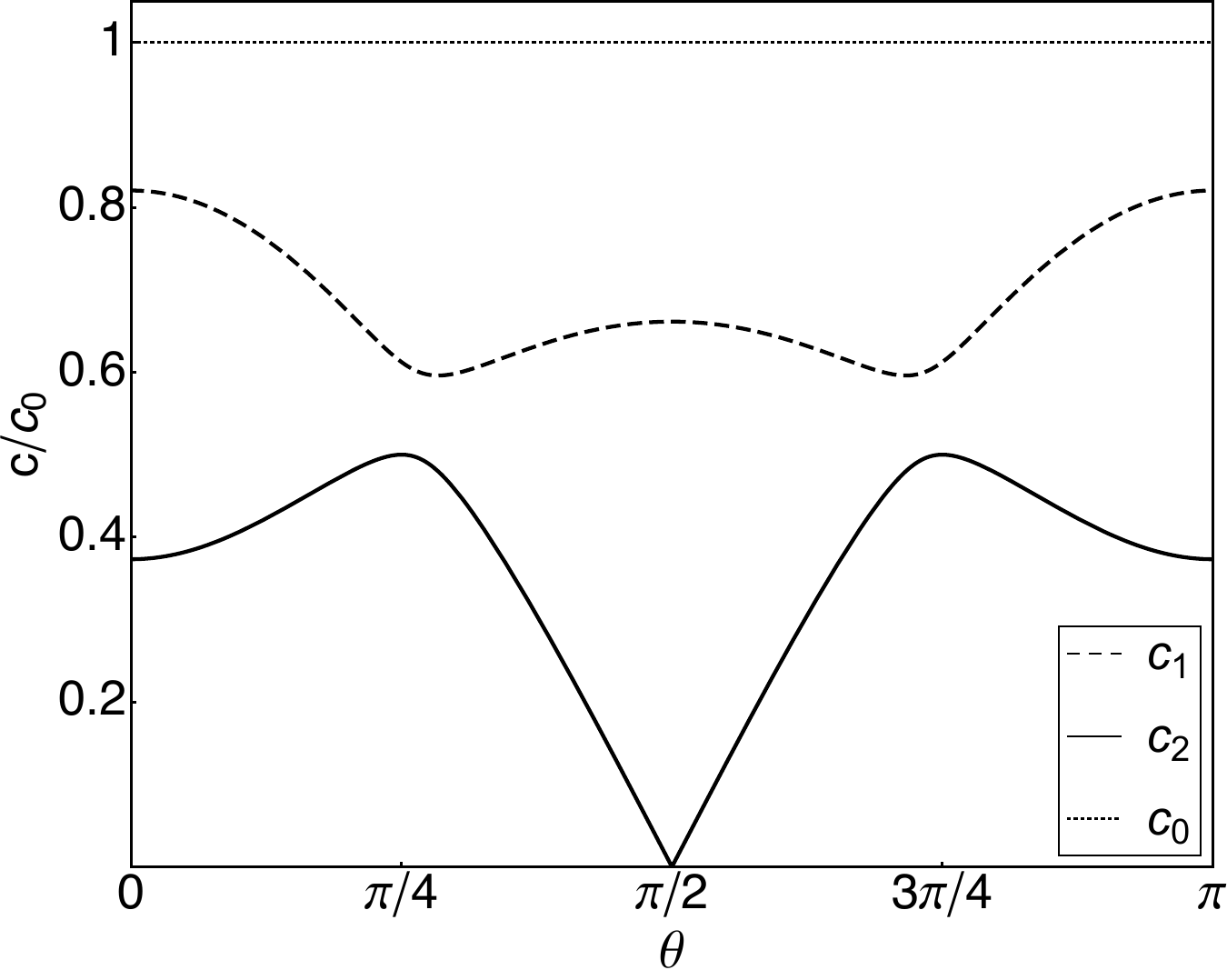}
\caption{Sound velocites as a function of the angle $\theta$ between the wave vector ${\bf k}$ and ${\bf p}_0$. There are two branches of sound excitations. The sound velocity of the lower branch (solid line) vanishes when ${\bf k}\perp {\bf p}_0$. The upper branch of sound excitation always has a finite velocity. The sound velocities are symmetric with respect to $\theta=\pi/2$ as it should be. $c_0$ is the sound velocity in the absence of $p$-wave attractive interaction. }
\label{fig:soundvelocity}
\end{figure}

To find the excitation spectrum, we write the small derivation from the ground state when ${\bf p}={\bf p}_0$ by
\begin{equation}
c_{\bf k}=\delta_{{\bf k},{\bf p}_0}+\delta c_{\bf k}~~,~d_{\bf k}=\delta_{{\bf k},-{\bf p}_0}+\delta d_{\bf k}
\end{equation}
and expand the function $F$ [defined below eq.(\ref{energy})] to second order in $\delta c_{\bf k}$ and $\delta d_{\bf k}$. The first order term vanishes by definition and one is left with
\begin{equation}
F=F_0+\sum_{\bf k}\Psi^\dagger_{\bf k}\mathcal{M}_{\bf k}\Psi_{\bf k}
\end{equation}
with $\Psi^\dagger_{\bf k}=(\delta c^*_{{\bf p}_0+{\bf k}}, \delta d^*_{-{\bf p}_0+{\bf k}}, \delta d_{-{\bf p}_0-{\bf k}}, \delta c_{{\bf p}_0-{\bf k}})$ and $\mathcal{M}$ is a four-by-four matrix whose matrix elements can be obtained analytically (see supplementary material). The excitation corresponds to the eigenvalue of the matrix $\gamma_5\mathcal{M}_{\bf k}$ with $\gamma_5$ being the standard Gamma matrix. It turned out that there are two elementary excitations $\omega_1({\bf k})$ and $\omega_2({\bf k})$, both tend to zero when $|{\bf k}|\to 0$. The analytic expression is fairly complicated and is given explicitly in the supplementary material. For general direction of ${\bf k}$ with respect to ${\bf p}_0$, the dispersion of $\omega_1({\bf k})$ and $\omega_2({\bf k})$ are linear and is characterised by two sound velocities
\begin{equation}
\omega_{1,2}({\bf k})=c_{1,2}({\hat{k}})|{\bf k}|
\end{equation}
where $\hat{k}$ denote a unit vector along ${\bf k}$. In Fig.(\ref{fig:soundvelocity}), we plot $c_{1,2}({\hat{k}})$ as a function of the angle $\theta$ between ${\bf k}$ and ${\bf p}_0$. Note first that because of the attractive $p$-wave interaction, the sound velocity is smaller than that for a purely $s$-wave system for which it is given by $c_0=\sqrt{nU_0/m}$. In particular, when $\theta=\pi/2$, the lower branch of the sound velocity $c_2({\hat{k}})$ vanishes when ${\bf k}\perp {\bf p}_0$. For this particular direction, 
\begin{equation}
\omega_2({\bf k})=\frac{1}{4m}\sqrt{36\pi^2a_1^2n^2-1}|{\bf k}|^2
\end{equation}
to lowest order in ${\bf k}$ and depends quadratically on $|{\bf k}|$. In addition, we note that the coefficient goes to zero at the transition point when a non-zero value of $|{\bf p}_0|$ develops [see eq.(\ref{valuep_0})], rendering an even softer sound mode at the onset of the fragmented condensate. Within our model, we have the standard simple condensate at ${\bf p}=0$ when $1+6\pi na_1>0$. The anisotropic sound velocities indicate an anisotropic superfluid behaviour that is particularly interesting in the direction perpendicular to ${\bf p}_0$.  An analogous situation occurs in the case of the spin-orbit coupled Bose gas at the transition point from zero momentum phase to the plane wave phase where the sound velocity vanishes~\cite{Ji2015}, rendering a zero superfluid density in the ground state for flow along the direction of spin-orbit coupling~\cite{Zhang2016, Martone2021}. In our case, the existence of a linear spectrum of $\omega_1({\bf k})$ for ${\bf k}\perp {\bf p}_0$ indicates that a finite superfluid density should still exist. 

{\it Spin Excitations.}--We now discuss the spin excitation of the system. Unlike the case for phonon wave functions in eq.(\ref{phonon}), for which the spin still remains a many-body singlet and the density fluctuation can be simply incorporated in the orbital states, one needs to go beyond the structure of wave function given in (\ref{phonon}). Since the Hamiltonian is $SU(2)$ invariant, it is clear that the spin excitation of the system can be labeled by the total spin ${\bf S}$ and its projection $S_z$, $\ket{S,S_z}$. Let us consider the lowest excitations that corresponds to ${\bf S}^2=2$ and $S_z=1,0,-1$. With standard notation, one can write the corresponding wave functions (not normalized) as,
\begin{align}
\ket{{1,0}} &=(a^\dagger_{0,\uparrow}a^\dagger_{1,\downarrow}+a^\dagger_{1,\uparrow}a^\dagger_{0,\downarrow})(A^\dagger)^{N/2-1}\ket{0}\\
\ket{{1,1}} &=a^\dagger_{0,\uparrow}a^\dagger_{1,\uparrow}(A^\dagger)^{N/2-1}\ket{0}\\
\ket{{1,-1}} &=a^\dagger_{0,\downarrow}a^\dagger_{1,\downarrow}(A^\dagger)^{N/2-1}\ket{0}
\end{align}
where we have defined the pair operator $A^\dagger=a^\dagger_{0,\uparrow}a^\dagger_{1,\downarrow}-a^\dagger_{1,\uparrow}a^\dagger_{0,\downarrow}$. It is easy to check that these three states are eigenstates of ${\bf S}^2$ and $S_z$, and furthermore are degenerate in energy. In fact, by a similar calculation as in the phonon case, one can show that for the $S=1$ multiplet, the ground state corresponds to $c_{\bf k}=\delta_{{\bf k},{\bf p}_0}$ and $d_{\bf k}=\delta_{{\bf k},-{\bf p}_0}$, the same as the phonon case. The energy can be calculated similarly. The kinetic energy remains the same and the interaction energy is instead given by
\begin{equation}
\frac{1}{V}\left(\frac{N^2}{2}-N+2\right)U_0+\frac{1}{V}\left(\frac{N^2}{4}+\frac{N}{2}-2\right)U_{{\bf p}_0,{\bf p}_0}^{\ell=1}
\end{equation}
Compare this with Eq.(\ref{energy_finite_N}), we find that the $s$-wave interaction energy is increased by $2U_0/V$ while the $p$-wave interaction is reduced by $2U_{{\bf p}_0,{\bf p}_0}^{\ell=1}/V$. Thus the excitation energies of states $\ket{1,\pm 1}$ and $\ket{1,0}$ are given by $2(U_0-U_{{\bf p}_0,{\bf p}_0}^{\ell=1})/V$. For higher spin excitations, the structure of the wave function becomes more complicated and will be left for future investigations.

{\em Conclusion}. We have shown that the inclusion of attractive $p$-wave attraction in an otherwise repulsive two-component Bose gas can lead to a fragmented condensate with singlet pairs playing an important role. The excitations of the system feature anisotropic sound velocities that vanishes in the direction perpendicular to ${\bf p}_0$ which characterizes the ground state. We expect that similar physics might also occur for higher partial wave channels, for example, in the recently realized $g$-wave molecular condensate~\cite{Zhang2021,Zhang2023}. In addition, our model gives rise to a quantum phase transition between a simple and a fragmented BEC. In the future, it will be of interest to investigate the topological excitations in the fragmented condensate and to characterize the phase transition in detail.

\paragraph{Acknowledgements.}
This work is supported by HK GRF Grants No. 17304820 and No. 17304719, CRF Grants No. C6009-20G and No. C7012-21G, and a RGC Fellowship Award No. HKU RFS2223-7S03.


\begin{thebibliography}{References}
\bibitem{SKMU} Dan M. Stamper-Kurn and Masahito Ueda, Rev. Mod. Phys. {\bf 85} 1191 (2013)
\bibitem{Choi2012} Choi, J. Y., W. J. Kwon, and Y. I. Shin, Phys. Rev. Lett. {\bf 108}, 035301 (2012).
\bibitem{Leslie2009} Leslie, L. S., A. Hansen, K. C. Wright, B. M. Deutsch, and N. P. Bigelow, Phys. Rev. Lett. {\bf 103}, 250401 (2009).
\bibitem{Jae-yoon2012} Jae-yoon, C., K. Woo Jin, L. Moonjoo, J. Hyunseok, A. Kyungwon, and S. Yong-il, New J. Phys. {\bf 14}, 053013 (2012)
\bibitem{Chang2005} Chang, M.-S., Q. Qin, W. Zhang, L. You, and M. Chapman, Nat. Phys. {\bf 1}, 111 (2005)
\bibitem{Kronjager2006} Kronjager, J., C. Becker, P. Navez, K. Bongs, and K. Sengstock, Phys. Rev. Lett. {\bf 97}, 110404 (2006).
\bibitem{Black2007} Black, A. T., E. Gomez, L. D. Turner, S. Jung, and P. D. Lett, Phys. Rev. Lett. {\bf 99}, 070403 (2007)
\bibitem{Jung2009} Liu, Y., S. Jung, S. E. Maxwell, L. D. Turner, E. Tiesinga, and P. D. Lett, Phys. Rev. Lett. {\bf 102}, 125301 (2009).
\bibitem{KS} A. B. Kuklov and B.V. Svistunov, Phys. Rev. Lett. {\bf 89}, 170403 (2002)
\bibitem{AL} S. Ashhab and A.J. Leggett, Phys. Rev. A {\bf 68}, 063612 (2003)
\bibitem{Leggettbook} A.J. Leggett, {\em Quantum Liquids}, Oxford University Press, 2006.
\bibitem{HammerLee} H.-W. Hammer and Dean Lee, Physics Letters B {\bf 681}, 500 (2009).
\bibitem{supp} In the supplementary material, we discuss the properties of the many-body ansatz eq.(\ref{ansatz1}) and calculate the expectation values of generic one-body and two-body operators. In particular, we calculate the ground state as well as the excited state energies and derive the analytic expressions for the velocities of sounds. We also implement the two-body scattering in an one-dimensional optical lattice and show how the $p$-wave effective range can be modified. 
\bibitem{LPB} C. K. Law, H. Pu, and N. P. Bigelow, Phys. Rev. Lett. {\bf 81}, 5257 (1998)
\bibitem{HoYip} Tin-Lun Ho and Sung Kit Yip, Phys. Rev. Lett. {\bf 84}, 4031 (2000)
\bibitem{RCPRL} L. Radzihovsky and S. Choi, Phys. Rev. Lett. {\bf 103}, 095302 (2009).
\bibitem{CRPRB} S. Choi and L. Radzihovsky, Phys. Rev. A {\bf 84}, 043612 (2011). 
\bibitem{Li2019} Zehan Li, Jian-Song Pan, and W. Vincent Liu, Phys. Rev. A. {\bf 100} 053620 (2019)
\bibitem{Castin2001} Y. Castin and C. Herzog, Comptes Rendus de l'Académie des Sciences - Series IV - Physics, {\bf 3}, 419 (2001)
\bibitem{Mueller2006} E. Mueller, Tin-Lun Ho, M. Ueda and G. Baym, Physical Review A {\bf 74}, 033612 (2006)
\bibitem{SLH} Sergio Lerma-Hern\'{a}ndez, Jorge Dukelsky, and Gerardo Ortiz, Research {\bf 1}, 032021(R) (2019)
\bibitem{Ji2015} S.-C. Ji, L. Zhang, X.-T. Xu, Z. Wu, Y. Deng, S. Chen, and J.-W. Pan, Phys. Rev. Lett. {\bf 114}, 105301 (2015).
\bibitem{Zhang2016} Y.-C. Zhang, Z.-Q. Yu, T. Kai Ng, S. Zhang, L. Pitaevskii and S. Stringari, Phys. Rev. A {\bf 94}, 033635 (2016) 
\bibitem{Martone2021} Giovanni Martone and Sandro Stringari, SciPost Phys. {\bf 11} 092 (2021)
\bibitem{Zhang2021} Zhengdong Zhang, Liangchao Chen, Kai-Xuan Yao and Cheng Chin, Nature {\bf 592}, 708 (2021)
\bibitem{Zhang2023} Zhengdong Zhang, Shu Nagata, Kai-Xuan Yao and Cheng Chin, Nature Physics {\bf 19}, 1466 (2023)

\end{thebibliography}
\end{document}